\documentclass[11pt]{article}
\usepackage[margin=1in]{geometry}

\usepackage{amsmath,amssymb,amsfonts}
\usepackage{mathtools}
\usepackage{amsthm}

\theoremstyle{remark}
\newtheorem{remark}{Remark}

\usepackage[round,authoryear]{natbib}
\usepackage{algorithm}
\usepackage{algpseudocode}

\numberwithin{equation}{section}

\newcommand{\holder}{H\"older }
\newcommand{\cadlag}{c\`adl\`ag }

\usepackage{booktabs}
\usepackage{xcolor}
\usepackage{pifont}

\DeclarePairedDelimiter{\norm}{\lVert}{\rVert}
\DeclarePairedDelimiter{\parenthesis}{(}{)}

\DeclarePairedDelimiter{\set}{\{}{\}}

\title{Higher-Order Efficient Estimators: A Review and Simulation-Based Benchmark Study}
\author{Zeyi Wang and Mark J. van der Laan}

\begin{document}

\maketitle

\begin{abstract}
Higher-order efficient estimators extend standard first-order semiparametric estimators by replacing second-order residuals with third- or higher-order terms, potentially enabling asymptotic efficiency under slower nuisance function convergence rates and improving finite-sample performance. Existing methods achieve higher-order expansions through structurally different approximation strategies, including basis truncation, kernel smoothing, and highly adaptive lasso (HAL) representations, making direct theoretical and practical comparison difficult. In this manuscript, we provide a focused review and a simulation-based empirical benchmark for second-order efficient estimators, using treatment-specific mean estimation as a canonical causal inference and missing data problem. We compare how higher-order influence function (HOIF) estimators, kernel-based higher-order targeted minimum loss-based estimator (HOTMLE), and HAL-based HOTMLE construct higher-order expansions and the approximation or regularization burdens they introduce. The asymptotic and numerical study evaluates first-order and empirical second-order estimators under controlled nuisance errors with constant or increasing sectional variation complexity. Results show that higher-order debiasing can substantially reduce first-order estimation bias; however, gains depend strongly on stability of the approximation or regularization required for higher-order correction. Empirical HAL-based HOTMLE shows relatively stable performance, while empirical HOIF remains sensitive to basis truncation and tuning choices. Overall, this manuscript clarifies when higher-order asymptotic improvements are attained in theory, when they may be practically visible, and when implementation instability may offset theoretical advantages.
\end{abstract}


\section{Introduction}

Targeted learning \citep{vanderLaan&Rubin06,vanderLaan&Rose11,vanderLaan&Rose18} is built around the core idea that flexible estimation of nuisance features of the data-generating distribution can be combined with semiparametric efficiency theory to produce estimators that are targeted for scientifically interpretable parameters. 
The two-step procedure separates potentially high-dimensional nuisance feature estimates from estimation of a low-dimensional target of inference. 
An initial nuisance estimate is updated along a parametric submodel whose score is determined by the efficient influence curve (EIC) of the target parameter. 
This targeting step is typically low-dimensional and likelihood-based, designed to solve the EIC equation, thereby controlling first-order bias for the target parameter while preserving structural constraints of the statistical model. The resulting targeted minimum loss-based estimator (TMLE) is a plug-in estimator, a property that can improve finite-sample stability and interpretability by respecting known bounds and model constraints.  

In causal inference and missing data problems, TMLEs have been widely used because they allow nuisance estimates to be estimated using flexible machine learning algorithms while retaining the large-sample guarantees of semiparametric efficient estimation \citep{van2007super,vanderLaan17}. They are closely related to other first-order efficient estimators, such as one-step estimators \citep{bkrw1997} and augmented inverse probability weighted estimators \citep{Robins1992RecoveryOI}, but differ in the plug-in construction and the use of targeted updates. These procedures are (first-order) asymptotically efficient \citep{Vaart_1998} in the sense that their asymptotic behavior is characterized by the EIC evaluated at the true distribution, through 
\begin{align}\label{eq:AL}
	\psi_n - \psi_0 = (P_n - P_0) D^*(P_0) + R_n + o_P(n^{-1/2}),  
\end{align}
which is typically achieved under regularity conditions involving Donsker type conditions or cross-fitting and a mild consistency condition with no rate requirement (e.g. Section \ref{sec:1TSM}). 
Here, $P_n$ is the empirical distribution, $P_0$ is the true distribution, $D^*$ is the EIC, and $R_n$ is a residual term, which in many doubly robust problems is second-order and bounded by a product of nuisance estimation errors in $L^2(P_0)$-norms. Asymptotic linearity and efficiency follow when $R_n = o_P(n^{-1/2})$. 
For example, in treatment-specific mean (TSM) estimation, or equivalently mean estimation under missing at random, the relevant nuisance functions are the outcome regression and the treatment or missingness propensity. If these nuisance components are estimated at comparable rates in $L^2(P_0)$ norm, then an $o_P(n^{-1/4})$ rate for each suffices for a diminishing residual on the root-n-scale. 
This second-order, product-rate residual structure is one of the main reasons first-order efficient estimators can successfully leverage flexible, data-adaptive nuisance estimation for improved bias control and valid root-n inference. 

The practicality of first-order efficient estimators has been substantially expanded by ensemble learning, cross-fitting, and highly adaptive regression. Super learner \citep{van2007super,Polley2025} and related ensemble implementations \citep{sl3Coyle2023,wang2025superensemblelearningusing} provide a principled way to combine multiple candidate algorithms and asymptotically achieve oracle risk performance relative to the candidate library under standard conditions. Cross-fitting weakens Donsker-type regularity conditions required for empirical process control. 
Highly adaptive lasso (HAL) \citep{vanderLaan15,Benkeser&vanderLaan16,vanderLaan17,Bibaut&vanderLaan19} guarantees nonparametric convergence rates approximately $O_P(n^{-1/3})$ (in $L^2(P_0)$-norm, ignoring $\log n$ factors) for \cadlag functions with bounded sectional variation norms. 
Therefore, first-order TMLE based on sufficiently flexible nuisance estimators, such as TMLE with initial estimates via super learners including HAL as a candidate algorithm (HAL-TMLE), generally achieves asymptotic efficiency in a broad range of causal and missing data problems, provided standard regularity and \cadlag nuisance functions with bounded sectional variation \citep{vanderLaan17}.

Nevertheless, first-order theory has key limitations. The product rate may fail, or may be only weakly satisfied in finite samples, when nuisance functions have high-frequency structure, complex interactions, or variation norms that grow with the sample size. Even when methods such as HAL-TMLE attain favorable asymptotic rates under conditions, finite-sample performance can still be affected by large variation constants or computational constraints. In these cases, first-order debiasing may leave a non-negligible residual asymptotically or at realistic sample sizes.

Higher-order estimation (HOE) was developed to address this limitation. The core idea is to move beyond first-order correction and construct estimators whose residual behaves like a third- or higher-order term. In favorable scenarios, such second- or higher-order correction can relaxe nuisance rate requirements, recover root-n behavior under slower nuisance convergence, or improve finite-sample bias. However, in general, (first-order) pathwise differentiability alone does not produce higher-order expansion even in canonical problems such as TSM estimation. Higher-order debiasing typically introduces new approximation or regularization errors and potentially nested bias-variance trade-offs. Thus, it is crucial to analyze not only when HOE improves asymptotic rates, but also when it yields stable empirical gains in finite samples. 

The main methodological families for higher-order estimation differ mainly in how they achieve higher-order expansions despite the lack of certain desired differentiability conditions. 
HOIF estimators \citep{robins2008imsnotes} circumvent a desired higher-order pathwise differentiability by utilizing higher-order U-statistic corrections after projection onto finite-dimensional bases motivated by \holder spaces. Empirical HOIF estimators \citep{liu2026semiparametricefficientempiricalhigher} replace challenging covariate density estimation with empirical Gram matrices. Kernel-based higher-order TMLE \citep{Caroneetal17} uses kernel smoothing to obtain approximate higher-order differentiability. HAL-based higher-order TMLE \citep{vanderlaan2021higherordertargetedmaximum} constructs data-adaptive, (first-order) pathwise-differentiable debiasing targets which also result in higher-order residuals. 
These methods aim for the similar ambition of higher-order residual behavior, but they rely on different structural assumptions and introduce distinct regularization burdens. 

A central difficulty is that higher-order debiasing is not free. The approximation used to enable higher-order expansions typically introduces new bias-variance trade-offs. For empirical HOIF estimators, one must choose a basis system, as well as a truncation dimension large enough to control projection bias but small enough to avoid destabilizing the U-statistic and the empirical Gram matrix inverses. Kernel-based higher-order TMLE replaces basis truncation with bandwidth selection, shifting additional regularization to smoothing bias and variance. HAL-based methods are based on milder differentiability conditions by exploiting HAL representation, but their performance still requires stability of sequential targeting steps and balance of representation error and estimation error. Thus, higher-order estimation changes the bias-variance problem; it does not eliminate it. 

This paper is motivated by the difficulty of directly comparing higher-order estimators and translating higher-order asymptotic results into practical guidance. 
The theory identifies general regimes where higher-order corrections should improve residual behavior, but it is often unclear how those regimes are related in practice, which regularization choices are required, and whether the predicted gains are visible at realistic sample sizes. 

We address these questions through a conceptual review of existing higher-order asymptotic results and a simulation-based empirical benchmark study. The review emphasizes the structure of higher-order expansions, the regularization burden induced by each estimator class, and nuisance settings in which asymptotic improvements are anticipated. 
The numerical study then evaluates whether these anticipated improvements are visible in finite samples under controlled data-generating processes designed to stress test different aspects of nuisance estimation and higher-order approximation. 

Our benchmark focuses on treatment-specific mean estimation, equivalently mean estimation under missing at random. This target is simple enough to allow transparent comparisons, but rich enough to illustrate main phenomena that motivate higher-order estimation. We compare first-order estimators with empirical higher-order procedures, emphasizing methods that avoid explicit covariate density estimation and require only moderate tuning, with the goal of preserving stability in practical implementation. The simulation design separates two often conflated challenges: the nuisance convergence rate and the difficulty of higher-order approximation. In particular, the benchmark includes variation controlled settings, where sectional variation is small, large, or increasing with sample sizes, and derivative-variation-controlled settings where oscillation complexity increases slower or faster. 

The contributions of the paper are three-fold. First, we provide a compact synthesis of higher-order estimation results most relevant for applied targeted learning, emphasizing practical and finite-sample implications. Second, we propose a benchmark framework that links simulation settings to different mechanisms where higher-order estimators are expected to succeed or fail, including nuisance convergence, regularization bias control, and numerical stability. Third, we report empirical evidence on when higher-order corrections appear useful or unstable, and identify implementation choices that appear important for practical use. Overall, this paper studies whether the asymptotic possibilities survive the approximation and regularization required for implementation. 

The remainder of the paper is organized as follows. Section 2 introduces the treatment-specific mean problem and first-order efficiency concepts. Section 3 reviews empirical higher-order estimators with emphasis on their regularization burdens and asymptotic implications. Section 4 presents the benchmark framework, including the five nuisance complexity settings and evaluation metrics. Section 5 reports simulation results. Section 6 discusses practical lessons, limitations, and open problems for higher-order targeted learning.

\section{Background and Preliminaries}

\subsection{Notations for Treatment-Specific Mean (TSM) Example}

Observed data is $O = (X, A, Y)$, where $X$ is the $d$-dimensional baseline covariates, $A$ is the binary treatment (1 for treated), and $Y$ is the outcome. Under standard causal assumptions (consistency, conditional ignorability, positivity), outcome regression $b(x) = E[Y | A = 1, X = x]$ identifies $E[Y(1) | X = x]$, where $Y(1)$ is the counterfactual outcome under $A=1$. The target of inference is then $\psi = E[b(X)]$, the treatment-specific mean (TSM). 
\begin{align*}
	b(x) &= E[Y | A = 1, X = x] \\
	g(x) &= P(A = 1 | X = x), \\
	f_X(x) &= \frac{dF_X}{d\mu}(x) \\
	\psi &= E[b(X)]. 
\end{align*}
Components $p = (b, g, f_X)$ for outcome regression, propensity, and covariate density factorize the full data distribution $P \in \mathcal{M}$ for binary outcomes, or the information of $P$ relevant to $\psi$ for continuous outcomes. With empirical higher-order estimators, the covariate density $f_X$ is estimated by the empirical measure, $P_n$, that puts equal weights to IID observations.

\subsection{First-Order Efficient Estimators for TSM}\label{sec:1TSM}

The TSM parameter $\psi = \Psi(P)$ is pathwise differentiable with efficient influence curve (EIC) $D^*(P) = \frac{A}{g(X)}(Y - b(X)) + b(X) - \Psi(P)$. In standard first-order estimation, $f_X$ is estimated with empirical distribution $P_n$, and the resulting estimator depends on $P$ only through outcome regression $b$. With a slight abuse of notation, we write $\Psi(P) = \Psi(b)$. For the true parameter value $\psi_0 = \Psi(b_0)$ at $P_0$, the second-order residual is $R(P, P_0) \equiv \Psi(P) - \Psi(P_0) + P_0 D^*(P) = P_0 \frac{(g-g_0)(b-b_0)}{g} $, which yields an exact expansion
\begin{align}
	\Psi(P) - \Psi(P_0) &= - P_0D^*(P) + R(P, P_0) \label{eq:onestep}\\
	&= (P_n - P_0)D^*(P) - P_n D^*(P) + R(P, P_0) \label{eq:1expansion}.
\end{align}
For an estimator $P$ of $P_0$ based on $n$ i.i.d. observations, under regularity conditions that $D^*(P)$ falls into a $P_0$-Donsker class with probability tending to $1$, $||D^*(P) - D^*(P_0)||_{L^2(P_0)} = o_P(1)$, $P_n D^*(P) = o_P(n^{-1/2})$, it achieves 
$$(P_n - P_0)\parenthesis{D^*(P) - D^*(P_0)}  = o_P(n^{-1/2})$$
and thereby the desired expansion (\ref{eq:AL}) with $R_n = R(P, P_0)$, which is bounded by $||g - g_0||_{L^2(P_0)}||b - b_0||_{L^2(P_0)}$ under positivity. Therefore, asymptotic linearity and efficiency of first-order estimators depend mainly on a product nuisance rate of $n^{-1/2}$. This is satisfied, for example, with HAL-based estimation for \cadlag, variation bounded functions, but still motivates third- or higher-order product nuisance rates for finite-sample performance, unbounded variation, and computational challenges. 

\begin{remark}\label{re:1regularity}
The Donsker class condition may be replaced by cross-fitting $b, g$ using oracle inequality. First-order TMLE achieves $P_n D^*(P^*) = o_P(n^{-1/2})$ at the targeted update $P^* = P^{(1)}_n(P) = P^{(1)}(P, \epsilon^*)$ by solving an MLE $\epsilon^* = \arg\max_\epsilon P_n \log p^{(1)}(P, \epsilon)$ along the least favorable parametric submodel $P^{(1)}(P, \epsilon)$ with score $\frac{d}{d\epsilon}_{|\epsilon = 0} \log p^{(1)}(P, \epsilon)=\frac{A}{g(X)}(Y - b(X))$. In TSM with binary or bounded continuous outcomes, it is typically chosen as $b^{(1)}(b, \epsilon) = \text{expit}\{\text{logit} b + \epsilon \frac{A}{g}\}$ where $H = \frac{A}{g}$ is the clever covariate in the targeting step logistic regression. The update $P^*$ preserves the nuisance rate in (\ref{eq:1expansion}) through the local parametric MLE $\epsilon^*$ under positivity so that $||b^* - b_0||_{L^2(P_0)} = O_P(||b - b_0||_{L^2(P_0)})$. 
$P_n D^*(P^*) = o_P(n^{-1/2})$ is not required for one-step estimator (the AIPW estimator in TSM) $\psi^* = \Psi(P) + P_n D^*(P)$, which alters (\ref{eq:onestep}) directly. $||D^*(P) - D^*(P_0)||_{L^2(P_0)} = o_P(1)$ is generally a weaker condition than $ ||g - g_0||_{L^2(P_0)}||b - b_0|||_{L^2(P_0)} = o_P(n^{-1/2})$. 	
\end{remark}

\section{Review of Empirical Higher-Order Efficient Estimators}

\subsection{Higher-Order Efficiency}

Second-order efficient estimators typically reach an expansion of the following form under additional differentiability conditions: 
\begin{align}\label{eq:HO}
	\psi_n - \psi_0 = (P_n - P_0) D^{(1)}(P_0) + \mathbb{S}_n^{(2)} + E_n + R_n  + o_P(n^{-1/2}), 
\end{align}
where $D^{(1)}$ represents the first-order EIC, $\mathbb{S}_n^{(2)}$ denotes a second-order debiasing term, often a U-statistic or higher-order empirical process term, $E_n$ is an additional approximation or representation error, and $R_n$ is typically bounded by third-order products of nuisance estimation errors. 
For asymptotic linearity, $\mathbb{S}_n^{(2)} = o_P(n^{-1/2})$ is needed, but the required conditions may be more challenging if (\ref{eq:HO}) relies on more demanding differentiability conditions. 
%which can often be achieved through degeneracy combined with appropriate rates for the approximation dimension $k$ or the kernel bandwidth $h$. 
%Under additional Donsker type conditions or cross-fitting and consistency (similar to Section \ref{sec:1TSM}), $\mathbb{S}_n^{(2)} = o_P(n^{-1/2})$ can be achieved 
Overall, compared with the first-order expansion (\ref{eq:AL}), the second-order residual is replaced by a higher-order residual, which may allow semiparametric efficiency under slower nuisance convergence rates, provided proper control of the induced approximation error $E_n$. 
For concrete and traceable comparisons, this manuscript primarily focuses on second-order estimators, although higher-order generalizations are possible. 

However, the existence of expansion (\ref{eq:HO}) is non-trivial in causal inference and missing data problems. In such settings, the original target parameter typically fails to reach the required differentiability, and practical higher-order estimation often relies on an approximated representation, which induces the additional error $E_n$ that does not appear in the first-order expansion (\ref{eq:AL}). 

This distinction leads to a crucial yet delicate bias-variance trade-off. The third-order residual control is dominant only after reducing the smoothing, truncation, or approximation error $E_n$. 
For example, reducing smoothing decreases $E_n$ and improves the approximation, but at the same time may approach practical near-violation of the required pathwise-differentiability. As a result, $\mathbb{S}^2_n$ may explode in finite samples even when asymptotic linearity formally holds. Thus, it is important to evaluate how the higher-order expansion (\ref{eq:HO}) is constructed and how demanding the approximate differentiability conditions are in practice. In the rest of this section, we review two classes of higher-order constructions that face distinct differentiability challenges.

\subsection{Constructions Based on Second-Order Canonical Gradient}

When second-order pathwise differentiability holds, the exact expansion (\ref{eq:onestep}) can be refined by approximating the second-order residual through a second-order canonical gradient: 
\begin{align*}
	\Psi(P) - \Psi(P_0) &= - P_0D^{(1)}(P) - \frac{1}{2}P_0^2D^{(2)}(P) + R_n \\
	&= (P_n- P_0)D^{(1)}(P) + \frac{1}{2} (P_n^2 - P_0^2)D^{(2)}(P) - P_n D^{(1)}(P) - \frac{1}{2} P_n^2 D^{(2)}(P) + R_n, 
\end{align*}
where $- \frac{1}{2}P_0^2D^{(2)}(P)$ approximates the second-order remainder up to a third-order error using the second-order canonical gradient, $D^{(2)}(P)$. Here, we adopt the notation $P_0^2 h = \iint h(o_1, o_2) dP_0(o_1)dP_0(o_2)$ for expectation of bivariate maps over two independent draws, and $P_n^2h$ results in the V-statistic. 
This expansion motivates a class of second- or higher-order efficient estimators, designed as one-step estimators or TMLEs, that essentially depend on the existence of higher-order canonical gradients. Additional Donsker type conditions are required for higher-order inference utilizing both first- and second-order canonical gradients, but are not required for attaining asymptotic linearity. 

Unfortunately, second-order pathwise differentiability may not hold even for canonical target parameters such as TSM. Given $\Psi(P) = E[b(X)] = \int b(x) f_X(x) dx$ where $b(x) = E[\frac{\delta(X - x)}{f_X(X)}b(X)] = E[\frac{\delta(X - x)}{f_X(X)}\frac{A}{g(X)}Y]$ and $f_X(x) = E[\delta(X - x)f_X(X)]$, we derive the formal second-order canonical gradient with independent copies $O_1, O_2$ 
\begin{align*}
%	D^{(1)}(P)(O) &= \frac{A}{g(X)}(Y - b(X)) + b(X) - \Psi(P) \\
	D^{(2)}(P)(O_1, O_2) &= \frac{2A_1}{ g(X_1)}
	\left\{Y_1 - b(X_1)\right\}
	\frac{\delta(X_1 - X_2)}{f_X(X_1)}
	\left\{1-\frac{A_2}{g(X_2)}\right\}, 
\end{align*}
which involves Dirac's delta and is not square-integrable for continuous $X$. As a result, the TSM parameter is not genuinely second-order pathwise differentiable. 

To solve this non-differentiability issue, the following methods adopt two different approximations: the HOIF-based method approximates $\frac{\delta(X_1 - X_2)}{f_X(X_1)g(X_1)} \approx \bar z_k(X_1)^{\top}\Omega^{-1}\bar z_k(X_2)$ with a finite-dimensional projection, whereas the kernel-based HOTMLE applies kernel-smoothing for $\delta(X_1 - X_2)$. 
Such truncation or smoothing produces an approximated version of the second-order residual, which can then be estimated with a third-order error, leading to a combined error of $E_n + R_n$ due to approximation (truncation or smoothing) and estimation. 

Although not analyzed in these publications, they also correspond to estimating truncated or smoothed auxiliary parameters $$\Psi_{\bar z_k}(P) = E[b_{\bar z_k}(X)], \qquad \Psi_{h}(P) = E[b_{h}(X)],$$ where $$b_{\bar z_k}(x) = E[\bar z_k(X)^{\top}\Omega^{-1}\bar z_k(x) AY]$$ and $$b_h(x) = E[\frac{K_h(X - x)}{f_X(X)}b(X)]$$ are dependent on the choice of basis $\bar z_k$ and the kernel $K_h$. 
In addition, since $E(AY | X) = g(X)b(X)$, we have $b_{\bar z}(x) = \bar z_k(x)^\top\Omega^{-1}E[g(X)\bar z_k(X)b(X)] = \Pi^{(g)}_{\bar z_k}b(x)$; that is,  $b_{\bar z_k}$ is the orthogonal projection of $b$ onto span$\{\bar z_k\}$ in the weighted Hilbert space $L^2(gf_X)$, where the projection is defined under the inner product $\langle u, s \rangle_g = E[g(X)u(X)s(X)]$ with the Gram matrix $\Omega = E[g(X)\bar z_k(X)\bar z_k(X)^\top]$. Equivalently, $$b_{\bar z_k} = \arg\min_{s\in\mathrm{span}\{\bar z_k\}}E[g(X)\{b(X) - s(X)\}^2],$$ so that $b_{\bar z_k}$ can be viewed as the weighted least-square regression of $b(X)$ onto $\bar z_k(X)$ with weights $g(X)$. 

Both approaches result in a third-order remainder $R_n$ that involves nuisance errors for $b, g, f_X$ and an approximation error $E_n$ due to truncation or smoothing. We review the detailed constructions below. 

\subsubsection{HOIF}

HOIF-based second-order estimator relies on the following second-order U-statistic
%, degenerated at $P_0$, 
for additional debiasing: 
\begin{align}
\mathbb{S}_n^{(2)} &= \frac{1}{n(n-1)} \sum_{i_1 \not= i_2} -A_{i_1}(Y_{i_1}-b(X_{i_1}))\bar z_k(X_{i_1})^{\top}\Omega^{-1}\bar z_k(X_{i_2})(1- \frac{A_{i_2}}{g(X_{i_2})}), \label{eq:hoif}
\end{align}
where $\bar z_k(x) = (z_1(x), \dots, z_k(x))^T$ is a chosen basis system with $k$ dimensions, and $\Omega = P_0[A \bar z_k(X)\bar z_k(X)^T] $ is the Gram matrix. 
(\ref{eq:hoif}) is a standard U-statistic correction based on estimating the first-order estimator's cross-product residual, $R(P, P_0) = \int (b(x)-b_0(x))(\frac{1}{g(x)} - \frac{1}{g_0(x)})g_0(x) f_{X0}(x) dx = \langle b(x)-b_0(x), \frac{1}{g(x)} - \frac{1}{g_0(x)}\rangle_{g_0} \approx \left\langle \Pi^{(g)}_{\bar z_k}(b(x)-b_0(x)), \Pi^{(g)}_{\bar z_k}(\frac{1}{g(x)} - \frac{1}{g_0(x)})\right\rangle_{g_0} = E_0[g_0 (b-b_0) \bar z_k ]^{\top} \Omega^{-1} E_0[g_0 (\frac{1}{g}-\frac{1}{g_0}) \bar z_k ] $ after 
projecting the nuisance-error functions
$b-b_0$ and $\frac{1}{g} - \frac{1}{g_0}$ onto the functional space spanned by $\bar z_k(x)$. Therefore, the error of estimating the second-order residual is decomposed into the truncation bias $E_n$ due to projections and the third-order error $R_n$ of estimating the projected version with the U-statistic. 
%In addition, with the bilinear $D^{(2)}$ degenerate at $P_0$ (e.g. in TSM), $\mathbb{S}^{(2)}_n = - \frac{1}{2}P_n^2D^{(2)}(P)$
% $- \frac{1}{2}P_0^2D^{(2)}(P)$ is naturally approximated by the U-statistic . 
This results in a second-order HOIF estimator of the form $\psi_n = \Psi(P) + P_n D^{(1)}(P) + \mathbb{S}_n^{(2)}$. 

Formally, the second-order U-statistic involves a Dirac delta $\delta(x_1 - x_2)$ satisfying $\int \delta(x_1 - x_2) f(x_2) dx_2 = f(x_1)$. 
For discrete covariates with finite support, this is square-integrable and requires no projection approximation so that the truncation error $E_n$ vanishes. 
For continuous covariates, it is not square-integrable and does not yield a valid U-statistic. This motivates 
$\bar z_k(X_{i_1})^{\top}\Omega^{-1}\bar z_k(X_{i_2})$, which gives a square-integrable approximation to the formal kernel, and it makes the projected version of the second-order residual estimable by a U-statistic. 
The rest of the U-statistic construction is standard: it uses two distinct observations to estimate a cross-product of two (projected) nuisance errors. 
If the basis is orthonormal in $L^2(g_0f_{X0})$, then $\Omega = I_k$; otherwise, the correction $\Omega_0^{-1}$ must be estimated in practice.

To complete the estimator construction, it still requires specifying the basis system and estimating the Gram matrix. The original, density-based version uses a nonparametric density estimate $\hat f_X$; however, reliable density estimate can be challenging in practice. The empirical version uses the empirical covariance $\hat \Omega = P_n[A \bar z_k(X) \bar z_k(X)^T]$; in this case, it requires a slowly increasing $k$ that avoids exploding the inverse but still controls the truncation bias $E_n$. Although different bases may yield the same minimax rate, finite-sample performance can differ substantially depending on the actual data-generating process and the stability of $\hat \Omega^{-1}$. 

As the oracle U-statistic removes the projected component of the first-order estimator's residual, the leftover error consists of two components. 
The ``truncation bias'' arises from the basis approximation that enables the U-statistic construction, 
\begin{align*}
	E_n
	&=
	\int
	\{I-\Pi^{(g)}_k\}( b-b_0)(x)
	\{I-\Pi^{(g)}_k\}( 1/g-1/g_0)(x)
	g_0(x)f_{X0}(x)\,dx \\
	&= O_P(|| \{I-\Pi^{(g)}_k\}(b-b_0) ||_{L^2(g_0f_{X0})}  ||\{I-\Pi^{(g)}_k\}(1/g -1/g_0)||_{L^2(g_0f_{X0})})
\end{align*}
where $\Pi^{(g)}_k$ denotes the \(L^2(g_0f_{X0})\) projection onto the span of \(\bar z_k\). 
The ``estimation bias'' arises from estimating $\Omega$ and leads to a third-order term. With original HOIF estimators, 
\begin{align*}
	R_n
	=
	O_P\!\left(
	\| b-b_0\|_{L^2(P_0)}
	\| g-g_0\|_{L^2(P_0)}
	\| f_X-f_{X0}\|_{L^2(P_0)}
	\right),
\end{align*}
or with empirical HOIF estimators, 
\begin{align*}
	R_n
	=
	O_P\!\left(
	\| b-b_0\|_{L^2(P_0)}
	\| g-g_0\|_{L^2(P_0)}
	\|\hat\Omega-\Omega_0\|_{\mathrm{op}}
	\right). 
\end{align*}

Under \holder smoothness assumptions with coefficients $\beta_b, \beta_g, \beta_f$ for $b, g, f_X$ and bases satisfying Jackson approximation properties (e.g. wavelets, splines, Fourier), typically we have 
\begin{align*}
	E_n=O_P\!\left(k^{-(\beta_b+\beta_g)/d}\right). 
\end{align*}
Furthermore, for original, density-based HOIF, 
\begin{align*}
	R_n = O_P(n^{-\frac{\beta_b}{2\beta_b + d} - \frac{\beta_g}{2\beta_g + d} -\frac{\beta_f}{2\beta_f + d}}). 
\end{align*}
For empirical HOIF, 
\begin{align*}
	R_n = O_P(n^{-\frac{\beta_b}{2\beta_b + d} - \frac{\beta_g}{2\beta_g + d} } \sqrt{k/n}). 
\end{align*}
In addition, the U-statistic behavior will be dependent on the extent of truncation through $$\mathbb{S}^{(2)}_n = O_P(\frac{\sqrt{k}}{n}),$$ and therefore asymptotic linearity requires $k = o(n)$. 

The decomposition and the rate results highlight the central trade-off in HOIF methods: increasing $k$ is necessary to reduce approximation bias and improve representation accuracy, but may simultaneously induce statistical and numerical instability. For empirical HOIF, this regularization trade-off is particularly delicate: increasing $k$ may worsen both the U-statistic performance and the numerical stability of $\hat \Omega^{-1}$. 
In practice, large $k$ may lead to practical near-violation of second-order pathwise-differentiability and result in non-negligible constant in finite samples. 
At the same time, larger $k$ destabilizes inversion of the empirical Gram matrix; stable inversion at large $k$ requires larger $n$ and operator error $\|\hat\Omega - \Omega\|_{\text{op}}$ sufficiently small relative to the minimum eigenvalue of $\Omega$. Existing asymptotic results suggest carefully controlled growth rates for $k$ such as $k \sim n/(\log n)^2$, but finite-sample performance may remain highly sensitive to basis choice. In particular, some basis systems may require larger approximation dimensions to achieve adequate representation bias reduction, while simultaneously yielding unstable empirical behavior at the available sample size. 

\subsubsection{Kernel-based HOTMLE}

Kernel-based higher-order TMLE (HOTMLE) is motivated by the same challenge that leads to the finite-dimensional projection in HOIF: for continuous covariates, the formal second-order gradient involves a Dirac delta, and therefore does not define a square-integrable second-order influence function. Instead of replacing the delta with a projection kernel inspired by basis approximation, kernel-based HOTMLE replaces it with a smoothing kernel, $K(\frac{x_1-x_2}{h})$ indexed by a bandwidth $h$. Such kernel smoothing also yields an approximate second-order canonical gradient. The debiasing is then implemented through targeting steps within the TMLE framework, instead of as one-step updates. 

For the counterfactual mean example, this defines the following approximated second-order efficient influence curve: 
\[
D_h^{(2)}(P)(o_1,o_2)
=
H_h^{(2)}(P)(x_1,a_1,x_2,a_2)
\{y_1- b(x_1)\},
\]
with
\[
H_h^{(2)}(P)(x_1,a_1,x_2,a_2)
=
\frac{2a_1}{ g(x_1)f_X(x_1)}
K_h(x_1-x_2)
\left\{1-\frac{a_2}{g(x_2)}\right\},
\qquad
K_h(u)=h^{-d}K(u/h).
\]
The targeting step is then augmented so that the final targeted update \(P^*\) solves both
\[
P_n D^{(1)}(P^*)=0,
\qquad
P_n\bar D_{h_n,n}^{(2)}(P_n^*)=0
,
\]
where
$
\bar D_{h,n}^{(2)}(P)(o)
=
\frac{1}{n}\sum_{j=1}^n
D_h^{(2)}(P)(o,O_j)$, 
heuristically equivalent to solving $P_n^2 D_{h_n}^{(2)}(P^*)=0$. 

The resulting estimator is subject to the expansion \ref{eq:HO}, where 
\begin{align*}
	\mathbb{S}^{(2)}_n &= \frac{1}{2}(P_n^2-P_0^2)D_{h_n}^{(2)}(P) = O_P\!\left(\frac{h_n^{-d/2}}{n}\right) \\
%	E_n
%	&=
%	\frac{1}{2}
%	\left[
%	P_0^2D_{h_n}^{(2)}(P)
%	-
%	\lim_{h\to 0}P_0^2D_h^{(2)}(P)
%	\right]\\
	R_n
	&=
	O_P\parenthesis*{
	\norm*{
	1-\frac{f_{X0}g_0}
	{f_X g}	
	}_{L^2(P_0)}
	\norm*{
	\frac{g-g_0}{g}
	}_{L^2(P_0)}
	\norm*{
	b-b_0
	}_{L^2(P_0)}
	}, 
\end{align*}
and the approximation error $E_n$ due to smoothing is controlled as $E_n = O_P\!\left(
h_n^{m_0+1}
\| b-b_0\|_{L^2(P_0)}
\right)$ when $b_0, g_0$ are assumed $m_0$-times continuously differentiable and $K$ is $2d$-times differentiable. Asymptotic linearity requires selecting $h_n^{-d}=o(n)$.

Therefore, careful tuning is required for balancing the nuisance rates, assumed smoothness order, kernel choice, bandwidth, and precision of the kernel approximation. Since kernel HOTMLE replaces the HOIF basis-dimension trade-off with a bandwidth trade-off, its finite-sample performance likewise remains sensitive to precision of the density estimate $f_X$ through both approximation error $E_n$ and residual $R_n$. 

\subsection{Constructions Based on HAL and Fluctuation Parameters}

Constructions so far are motivated by estimating and correcting the second-order residual term $R_n$ in (\ref{eq:AL}) with approximations of the non-existent formal second-order canonical gradient. A closely related but distinct motivation is to directly debias a first-order estimator. Since $P_0 D^*(P_0) = 0$, one can consider minimizing 
\begin{align*}
	f_{n, 0}(P) = \{ \Psi(P^*(P)) - \Psi(P_0) - P_n D^*(P_0) \}^2, 
\end{align*}
where $P^*(P)$ is the first-order TMLE update of an initial estimator $P$. Along a regular parametric submodel $P_{\delta, h}$ through $P$ with score $h$, the pathwise derivative has the form 
\begin{align*}
	\frac{d}{d\delta}_{|\delta = 0} f_{n, 0}(P_{\delta, h}) = C(P) \cdot \frac{d}{d\delta}_{|\delta = 0}\Psi(P^*(P_{\delta, h})), 
\end{align*}
for a constant $C(P)$ conditioning on the initial estimator. This motivates targeting the fluctuation parameter $P \mapsto \Psi(P^*(P))$ directly. 

For continuous covariates, this fluctuation parameter is generally (first-order) non-pathwise differentiable, because $dP_n/dP$ does not exist for continuous $P$. 
This first-order non-differentiability issue is less challenging than the lack of second-order pathwise differentiability of the original parameter, since the differentiability condition demands only $o_P(n^{-1/2})$ precision (instead of $o_P(n^{-1})$) for the approximation of pathwise derivatives along regular parametric submodels. It is also reflected by the easier control of the debiasing term $\mathbb{S}^2_n$ as a standard empirical process under similar regularity conditions as first-order estimators.

HAL-based HOTMLE addresses this non-differentiability issue by replacing the empirical $P_n$ with a smooth surrogate $\tilde P_n$ from the highly adaptive lasso maximum likelihood estimator (HAL-MLE). The resulting fluctuation parameter becomes pathwise differentiable and admits a canonical gradient. 
Specifically, when an HAL-MLE $\Tilde P_n$ satisfies $d\Tilde P_n/dP \in L^2(P)$ and $\Psi: \mathcal{M} \to \mathbb{R}^k$ is pathwise differentiable, a data-adaptive, pathwise differentiable fluctuation parameter can be defined as 
\begin{align*}
	\Psi^{(1)}_n: P \mapsto \Psi(\Tilde P^{(1)}_n(P)), 
\end{align*}
where $\Tilde P^{(1)}_n(P)$ is a first-order targeted update using the HAL-MLE risk defined by $\tilde P_n$ instead of the empirical risk by $P_n$. The canonical gradient is denoted by $D^{(2)}_n$. 
Joint targeting of $\Psi(P_0)$ and the auxiliary parameter $\Psi^{(1)}_n(P_0)$ yields HAL-based second-order TMLE. 

\begin{remark}
	The original HAL-based HOTMLE is defined differently through sequential HAL-based updates,  $\Psi(\tilde P^{(1)}_n\tilde P^{(2)}_n(P))$, where the updates $\tilde P^{(1)}_n, \tilde P^{(2)}_n$ solve $\tilde P_n D^{(1)}(\tilde P^{(1)}_n(P)) = 0$, $\tilde P_n D^{(2)}_n(\tilde P^{(2)}_n(P)) = 0$. This version admits an expansion of the form 
	\begin{align*}
		\Psi(\tilde P^{(1)}_n\tilde P^{(2)}_n(P)) - \Psi(P_0) &= (P_n - P_0) D^{(1)}_{\tilde P^{(1)}_n(P)} + (\tilde P_n - P_0) D^{(2)}_{n, \tilde P^{(2)}_n(P)} + E_n + R_n + o_P(n^{-1/2}). 
	\end{align*}
	Under regularity conditions, asymptotic linearity can be attained with slower nuisance rates, as $(\tilde P_n - P_0) D^{(2)}_{n, \tilde P^{(2)}_n(P_0)}$ can be absorbed to a third-order residual $R_n$, and the additional regularization bias $E_n$ (due to the surrogate HAL-MLE) $(\tilde P_n - P_n) D^{(1)}_{\tilde P^{(1)}_n (P_0)}$ can be controlled through undersmoothing. 
\end{remark}

\begin{remark}
	Strictly speaking, only the definition of the fluctuation parameter $\Psi^{(1)}_n$ and its EIC $D^{(2)}_n$ depends on HAL-MLE, and the targeting steps can directly solve the empirical mean equations $ P_n D^{(1)}(P^{(1)}_n(P)) \approx 0$, $ P_n D^{(2)}_n(P^{(2)}_n(P)) \approx 0$. This empirical version, $\Psi( P^{(1)}_n P^{(2)}_n(P)) $,  leads to standard empirical process terms up to a third-order residual $R_n$ and a slightly different regularization bias $E_n$ led by $(\tilde P_n - P_n) \{ D^{(1)}_{P^{(1)}_n(P_0)} - D^{(1)}_{P^{(1)}_nP^{(2)}_n(P)} \}$. This may provide extra cancellation in the HAL regularization bias and may improve finite-sample stability. Throughout this manuscript, we focus on this empirical implementation of second-order TMLE. 
\end{remark}

This HAL-based second-order TMLE admits a higher-order expansion of the form  (\ref{eq:HO}) under regularity conditions of Section \ref{sec:1TSM}. Let $P^{(1)}_n(P), P^{(2)}_n(P)$ be the first- and second-order TMLE updates that solve $ P_n D^{(1)}(P^{(1)}_n(P)) = o_P(n^{-1/2})$ and $ P_n D^{(2)}_n(P^{(2)}_n(P)) = o_P(n^{-1/2})$. Then, 
\begin{align*}
	\mathbb{S}^{(2)}_n &= (P_n - P_0) D^{(2)}_{n, P^{(2)}_n(P)} \\
	E_n &= O_P\left((\tilde P_n - P_n) \{ D^{(1)}_{P^{(1)}_n(P_0)} - D^{(1)}_{P^{(1)}_nP^{(2)}_n(P)} \} + 
	(\tilde P_n - P_n) \{ D^{(2)}_{n, P^{(2)}_n(P_0)} - D^{(2)}_{n, P^{(2)}_n(P)}\}
	 \right) \\
%	R_n &= O_P\left(\|p - \tilde p_n\|_{L^2(P_0)}^3 + \|\tilde p_n - p_0\|_{L^2(P_0)}^3 \right). 
	R_n &= O_P\left(\|p - p_0\|_{L^2(P_0)}^3 + \|\tilde p_n - p_0\|_{L^2(P_0)}^3 \right). 
\end{align*}
Under additional Donsker conditions or cross-fitting and mild consistency similar to Section \ref{sec:1TSM}, generally we have $\mathbb{S}^{(2)}_n = (P_n - P_0) D^{(2)}_{n, P_0} + o_P(n^{-1/2}) = o_P(n^{-1/2})$, where $\norm{D^{(2)}_{n, P_0}}_{L^2(P_0)}$ involves first-order difference of $\tilde P_n, P_n$. 
%The regularization bias $E_n$ is determined by the difference of $\tilde P_n - P_n$, which is reduced when $\tilde P_n$ gets closer to $P_n$.

In the TSM example, this construction avoids estimating the covariate density. Similar to first-order estimators, the marginal covariate distribution $f_X$ is set to the marginal empirical distribution in both $\tilde P_n$ and $P$. Let $p = (b, g)$ be the estimated nuisance components. The residual satisfies 
\begin{align*}
	R_n = O_P(\|g - g_0\|\|p - p_0\|^2 + \|\tilde g_n - g_0\|\|\tilde p_n - p_0\|^2). 
\end{align*}
Under bounded sectional variation norm assumptions, or slowly increasing complexity that preserves $\|\tilde p_n - p_0\| = o_P(n^{-1/6})$, the residual is asymptotically third-order, represented by $R_n = O_P\left(\|g - g_0\|(\|g - g_0\|^2 + \|b - b_0\|^2)\right)$ in (\ref{eq:HO}). 

The approximation error $E_n$, in addition to the third-order residual, arises from replacing the empirical distribution $P_n$ by the HAL surrogate $\tilde P_n$. Unlike the truncation bias in HOIF or the smoothing bias in kernel-based HOTMLE, this regularization bias is controlled by the amount of HAL regularization and may be reduced through undersmoothing, that is, by allowing larger sectional variation norms than the cross-validated choice and allowing $\tilde P_n$ closer to $P_n$. As a result, it replaces the challenging basis or bandwidth tuning with tuning of sectional variation bounds, which motivates clearly defined finite-sample criteria of increasing the complexity bounds till $(\tilde P_n - P_n) D^{(1)}_{P^*} \approx 0$ or $(\tilde P_n - P_n) ( D^{(1)}_{P^{(1)}_n(P)} - D^{(1)}_{P^{(1)}_nP^{(2)}_n(P)} ) \approx 0$. 

The leading term $(\tilde P_n - P_n) \{ D^{(1)}_{P^{(1)}_n(P_0)} - D^{(1)}_{P^{(1)}_nP^{(2)}_n(P)} \}$ in the regularization bias $E_n$ is intuitively controlled in undersmoothing as $\tilde P_n \to P_n$, and two upper bounds can be applied. First, we have 
$$E_n = O_P(\norm{\tilde p_n - p_0}_{L^2(P_0)}\norm{p - p_0}_{L^2(P_0)})$$ 
directly from the leading term. Second, assuming $s \in \set{0, 1, \ldots}$-th order differentiability and sectional variation norm of the $s$-th order derivative is bounded by $K(n)$, 
$$E_n = O_P\parenthesis*{\norm{p-p_0}_v\norm{\tilde p_n - p_0}_{L^2(P_0)}\norm{\tilde u_{n, 0} - u_n}_{L^2(P_0)} }$$
where $\tilde u_{n, 0}$ is the projection of $u_n = \tilde v_n - \tilde P_n \tilde v_n$ onto the span of $J$ active basis in $\tilde p_n$, with $v_n = D^{(1)}_{P^{(1)}_n(P_0)} - D^{(1)}_{P^{(1)}_nP^{(2)}_n(P)}$ and $\tilde v_n = \frac{v_n}{\norm{v_n}_v}$. 
%We have $\norm{\tilde p_{n, 0} - p_0}_{L^2(P_0)} = O_P(\norm{\tilde p_{n} - p_0}_{L^2(P_0)})$ when $\tilde p_n$ is the cross-validated HAL-MLE and $K(n) = K<\infty$ is bounded. 
We have $\norm{\tilde u_{n, 0} - u_n}_{L^2(P_0)}$ of order $1/J$ and $\norm{\tilde p_{n} - p_0}_{L^2(P_0)}$ of order $\parenthesis*{(J/n)^{1/2} + (1/J)^{s+1}}K(n)$.  
%when $J$ is different from the cross-validated choice. 
When $s \geq 1$ and $p = \tilde p_n$, we have $\norm{\tilde p_n-p_0}_v$ of order $\parenthesis*{(J/n)^{1/2} + (1/J)^{s}}K(n)$. 
%Further derivation shows that the regularization bias $E_n$ is bounded by a product term of sectional variation norm complexity, an oracle approximation error as active basis approximating EIC, and an HAL approximation error. 
Therefore, we can have negligible $E_n = o_P(n^{-1/2})$ with or without undersmoothing in various settings under bounded or slowly increasing sectional variation norms, as illustrated in Section \ref{sec:rates}. 

Overall, the required conditions for asymptotic linearity are structurally different from \holder-type smoothness or kernel-based conditions. The fact that none or minimal undersmoothing is required for $E_n = o_P(n^{-1/2})$ leads to a minimal risk for near-violation of pathwise differentiability due to aggressively undersmoothing $\tilde P_n$. 
Even under moderate undersmoothing, $\mathbb{S}^2_n$ remains stable, because undersmoothing only selects larger variation bounds and typically preserves consistency and convergence rates of HAL. This again reflects the less demanding nature of first-order differentiability conditions regarding fluctuation parameters. 
The third-order residual $R_n$ is free of covariate density $f_X$, and now involves nuisance estimation errors from both an initial estimator $P$ and the HAL-MLE $\tilde P_n$.

\subsection{Empirical Estimators and Empirical Evidence}

Second-order estimators for TSM can be categorized by how they regularize the formal higher-order influence function and by whether covariate density estimation $f_X$ is required in addition to outcome and propensity regressions $b, g$. 
In practice, density estimation may be a bottleneck for scalability and introduce additional bias-variance trade-offs though bandwidth, basis, or smoothing hyperparameter selection. 
To clearly illustrate estimator performance under a stable, controlled benchmark framework, we focus on the empirical estimators that avoid explicit density estimation. This includes empirical HOIF using empirical Gram matrices and HAL-based HOTMLE solving empirical estimating equations, which are compared against standard first-order estimators. 

However, existing empirical evidence of these estimators are closely related to the underlying theoretical assumptions. For example, empirical analyses of HAL-based HOTMLE are conducted under bounded sectional variation settings, while HOIF analyses are typicallyevaluated under \holder-type smoothness conditions with basis satisfying classical approximation rates. As a result, the current empirical comparisons can be difficult to interpret across methods, analyzed under related but distinct data generating processes.

For example, minimax analysis characterizes the worst-case optimal rate over a specified function class, such as a \holder ball. But practical performance depends on the overlap between the true data-generating mechanism and the assumed function class. Realistic nuisance functions may have discontinuity, limited effective dimensions, or restricted oscillation patterns, which put them outside a nominal \holder class or within a strict submodel. Asymptotic minimax rates may then either overestimate or underestimate the nuisance rate in practice. Empirical evidence is lacking for many practically relevant settings, despite substantially available theoretical results.

Several important questions therefore remain only partially understood empirically:
\begin{itemize}
	\item When do higher-order estimators provide substantial finite-sample improvements over first-order estimators?
	\item When do higher-order estimators become unstable or provide only negligible gains? 
	\item How sensitive are different estimators to particular approximation or regularization choices? 
	\item To what extent do asymptotic results translate to practical finite-sample performance? 
\end{itemize}
A systematic empirical comparison across empirical higher-order estimators is needed for understanding the practical advantage and limitation of higher-order methods over standard first-order estimation.

\section{Benchmark Framework}

\subsection{Data Generating Process}

Covariate density $f_{X}$ is not directly estimated by empirical second-order estimators and first-order estimators. We stress test the stability of these empirical estimators by adopting a rough density function within a \holder class with coefficient $0.1$ and a complex covariance structure for a $4$-dimensional covariate $X \in [0, 1]^4$. 

Outcome and propensity score models $b, g$ are designed similarly, where $P(Y = 1 | A = 1, X) = b(X)$ and $P(A = 1 | X) = g(X)$ for binary $Y, A$. The truths $b_0, g_0$ are designed with controlled smoothness and complexity on the logit scale in the following five settings. 
\begin{enumerate}
	\item Small constant variation. 
	\item Large constant variation. 
	\item Increasing variation. 
	\item Slowly increasing oscillation (derivative variation). 
	\item Fast increasing oscillation (derivative variation). 
\end{enumerate}

Setting 1-3 are subsets of the Skorohod space $D[0, 1]^4$ of \cadlag functions, with a finite (0-order) bounded sectional variation norm $K$ or $K(n)$ given any finite sample size $n$. We denote these function classes as BV0 or BV0($K$) by the variation bound. Setting 3 allows $K(n)\to \infty$ and converges to the full $D[0, 1]^4$ as $n\to\infty$. \holder functions with coefficient $0 < \alpha < 1$, denoted by $H_\alpha[0, 1]^4$, may have infinite variation and each constructs a different subset of $D[0, 1]^4$: discontinuity is excluded by a \holder ball, infinite variation is excluded by BV0, and BV0 may contain functions satisfying \holder conditions up to $0 < \alpha \leq 1$. 

Setting 4-5 require almost everywhere differentiability and their (weak) derivative spaces contain \cadlag functions while converging to $D[0, 1]^4$. 
Except for allowing discontinuity, they construct subsets of the space of Lipschitz functions $H_1[0, 1]^4$ and therefore subsets of any \holder ball $H_\alpha[0, 1]^4$ with $0 < \alpha < 1$. We denote these smoother function classes as BV1 or BV1($K(n)$) with bounded first-order sectional variation norm $K(n)$ on the (weak) first-order derivatives. These can construct special nonparametric subspaces of \holder balls where further rate improvements may be possible, and they may expand outside the \holder balls through up to countably many discontinuity points. 

We use harmonic functions with increasing frequency and amplitude in Settings 1-5 for complexity control and they are defined over five equidistant sections per coordinate. In addition, we let Setting 1-3 also involve a mixture from $H_{1/4}[0, 1]^4$ with finitely many wavelet basis functions, so that Setting 1-3 DGPs are within $H_{1/4}[0, 1]^4$ except for finitely many discontinuity points between sections. Settings 4-5 are within $H_{1}[0, 1]^4$ except for discontinuity points. 
For improved traceability, we simulate with no-interaction so that effectively $d=1$ for Jackson-type minimax rates; however, 
the discontinuity may lead to Gibbs phenomenon and challenge the minimax rates. 

\subsection{Nuisance Rate Control}

We control the nuisance error rate exactly to the order of $n^{-1/6}$ such that second-order residual control is insufficient for root-n consistency and third-order residual control becomes necessary (except in the first-order competitors that use HAL instead as the initial estimators). 
Specifically, nuisance errors are controlled manually through $\hat b = b_0 + n^{-1/6} e_n$ with bounded $||e_n||_{L^2(P)}$, so that first-order estimates only result in a second-order residual of order $n^{-1/3}$, not diminishing asymptotically. Exact nuisance error control is implemented to stress-test residual-term performance, since minimax-based nuisance rates may be misleading in finite simulations. When the actual DGP lies in a strict submodel of the assumed \holder ball, minimax rates over the full ball can be over-pessimistic; this can occur when generating close to infinitely many basis functions is computationally expensive in simulations (e.g. Settings 1-3) or in realistic scenarios with potential rate improvements relative to the worst case (e.g. Settings 4-5). Minimax rates may also be over-optimistic when the \holder ball radius is poorly calibrated or when there exists discontinuity (e.g. Settings 1-5) so that the actual model expands outside the assumed class.

\subsection{Estimators}

We construct the following estimators using the initial estimation with controlled errors: first-order TMLE (1TMLE), empirical HAL-based second-order TMLE without or with additional undersmoothing (e2TMLE and u2TMLE), and empirical HOIF estimator (e2HOIF; see detailed basis construction in Section \ref{sec:rates}). Without further specification, these estimators share identical initial estimators, in order to illustrate the potential improvement of second-order estimators. 

Additionally, we compare the following first-order TMLE using HAL as the initial estimator: 0-order spline HAL as the initial (0HAL-1TMLE) in Setting 1-3 and 1-order spline HAL as the initial (1HAL-1TMLE) in Setting 4-5. These estimators are included as asymptotic state-of-the-art first-order competitors under each setting.

\subsection{Asymptotic Rates Comparison}\label{sec:rates}

We summarize below the combined asymptotic rates of $E_n + R_n$ in (\ref{eq:HO}) for second-order estimators or $R_n$ in (\ref{eq:AL}) for first-order estimators in Setting 1-5. 

For 1TMLE, $R_n$ is bounded by the second-order product of order $n^{-1/3}$ consisting of the nuisance estimation errors for $b, g$ of order $n^{-1/6}$. 
Similarly, for 0HAL-1TMLE, $R_n$ is bounded by the second-order product of 0HAL nuisance errors of order $n^{-1/3}$ (up to $\log n$ factors, same below in the subsection), which yields $n^{-2/3}$ with bounded sectional variation in Settings 1-2, and $n^{-4/9}$ in Setting 3 with increasing variation bounds $K(n) \sim n^{1/9}$; 1HAL attains $n^{-2/5}$ nuisance error rate under BV1, which yields 1HAL-1TMLE residuals of order $n^{-1/2}$ or $n^{-1/3}$ under BV1($K(n)$) with $K(n)\asymp n^{3/20}$ or $K(n)\asymp n^{7/30}$ in Settings 4-5. 

e2HOIF controls $R_n$ to the order of $n^{-5/6 + s/2}$ with $n^{-1/6}$ nuisance rates and $k \asymp n^{s}$ many Fourier basis functions, and leaves $E_n$ of order $n^{-s/2}$ or $n^{-2s}$ for $\beta = 1/4$ in Settings 1-3 or $\beta = 1$ in Settings 4-5 ignoring impact of discontinuity. The optimal combined rates for $E_n + R_n$ are up to $n^{-5/12}$ or $n^{-2/3}$ for Settings 1-3 or 4-5, attained at $k \asymp n^{5/6}/(\log n)^2$ or $n^{1/3}/(\log n)^2$, respectively. However, finite-sample implementation can be challenging, since $n^{5/6}/(\log n)^2$ ranges from 3 to 46, $n^{1/3}/(\log n)^2$ from 0.21 to 0.29, for $n \in [200, 25600]$, and there is no easy-to-follow tuning guidance. As an attempt to test the performance, we additively included $16$ Fourier basis per coordinate and the intercept, which led to $65 \times 65$ Fourier Gram matrices which typically still admit stable inversion. We hypothesized that $k = 65 < (25,600)^{1/2}$ may be sufficiently large for the approximation error up to discontinuity while not significantly deteriorating the residual order $n^{-5/6 + s/2}$.

Both e2TMLE and u2TMLE control $R_n = o_P(n^{-1/2})$ through products of nuisance estimation errors and the regularization bias $E_n$ may be of different orders. 
In Settings 1-3 with finite variation bound $K$ or slowly increasing $K(n)\asymp n^{1/9}$, 0HAL activates $J \asymp n^{1/3}$ basis, $\norm{p - p_0}_v$ is of order $n^{-1/6}$, and therefore $R_n$ dominates with or without undersmoothing. In Settings 4-5 under BV1($K(n)$) and increasing $K(n)$, 1HAL activates $J \asymp n^{1/5}$ basis, $\norm{\tilde p_n - p_0}_{L^2(P_0)}$ deteriorates linearly in $K(n)n^{-2/5}$, so we have $E_n$ at least of order $n^{-8/15}$ and thereby $R_n$ remains the dominant term before undersmoothing. 

\begin{table}[h!]
	\centering
	\begin{tabular}{lccc}
		\hline
		\textbf{Estimator} & \textbf{Setting 1--2} & \textbf{Setting 3} \\
		\hline
		1TMLE       & $O_P(n^{-1/3})$   & $O_P(n^{-1/3})$ \\
		e2TMLE      & $O_P(n^{-1/2})$   & $O_P(n^{-1/2})$ \\
		u2TMLE      & $O_P(n^{-1/2})$   & $O_P(n^{-1/2})$ \\
		e2HOIF      & $O_P(n^{-5/12})$  & $O_P(n^{-5/12})$ \\
		0HAL-1TMLE  & $O_P(n^{-2/3})$   & $O_P(n^{-4/9})$ \\
		\hline
	\end{tabular}
	\caption{Settings 1-2 correspond to BV0 with small or large constant variations, while Setting 3 corresponds to BV0($K(n)$) with increasing variation $K(n)\asymp n^{1/9}$. Nuisance rates are controlled to $n^{-1/6}$ except for 0HAL-1TMLE with HAL initial estimation. Rates are stated up to $\log n$ factors for the combined error $E_n + R_n$ of second-order estimators in (\ref{eq:HO}) or for the second-order residual $R_n$ of first-order estimators in (\ref{eq:AL}).}\label{tab:setting123}
\end{table}

\begin{table}[h!]
	\centering
	\begin{tabular}{lcc}
		\hline
		\textbf{Estimator} & \textbf{Setting 4} & \textbf{Setting 5} \\
		\hline
		1TMLE       & $O_P(n^{-1/3})$   & $O_P(n^{-1/3})$ \\
		e2TMLE      & $O_P(n^{-1/2})$   & $O_P(n^{-1/2})$ \\
		u2TMLE      & $O_P(n^{-1/2})$   & $O_P(n^{-1/2})$ \\
		e2HOIF      & $O_P(n^{-2/3})$   & $O_P(n^{-2/3})$ \\
		1HAL-1TMLE  & $O_P(n^{-1/2})$   & $O_P(n^{-1/3})$ \\
		\hline
	\end{tabular}
	\caption{Settings 4-5 correspond to BV1($K(n)$) with increasing variation;   $K(n)\asymp n^{3/20}$ in Setting 4 and $K(n)\asymp n^{7/30}$ in Setting 5. Nuisance rates are controlled to $n^{-1/6}$ except for 1HAL-1TMLE with  first-order spline HAL as initial estimation. Rates are stated up to $\log n$ factors for the combined error $E_n + R_n$ of second-order estimators in (\ref{eq:HO}) or for the second-order residual $R_n$ of first-order estimators in (\ref{eq:AL}).}\label{tab:setting45}
\end{table}

\section{Simulation Results}

\begin{figure}[t]
	\centering
	\includegraphics[width=\linewidth]{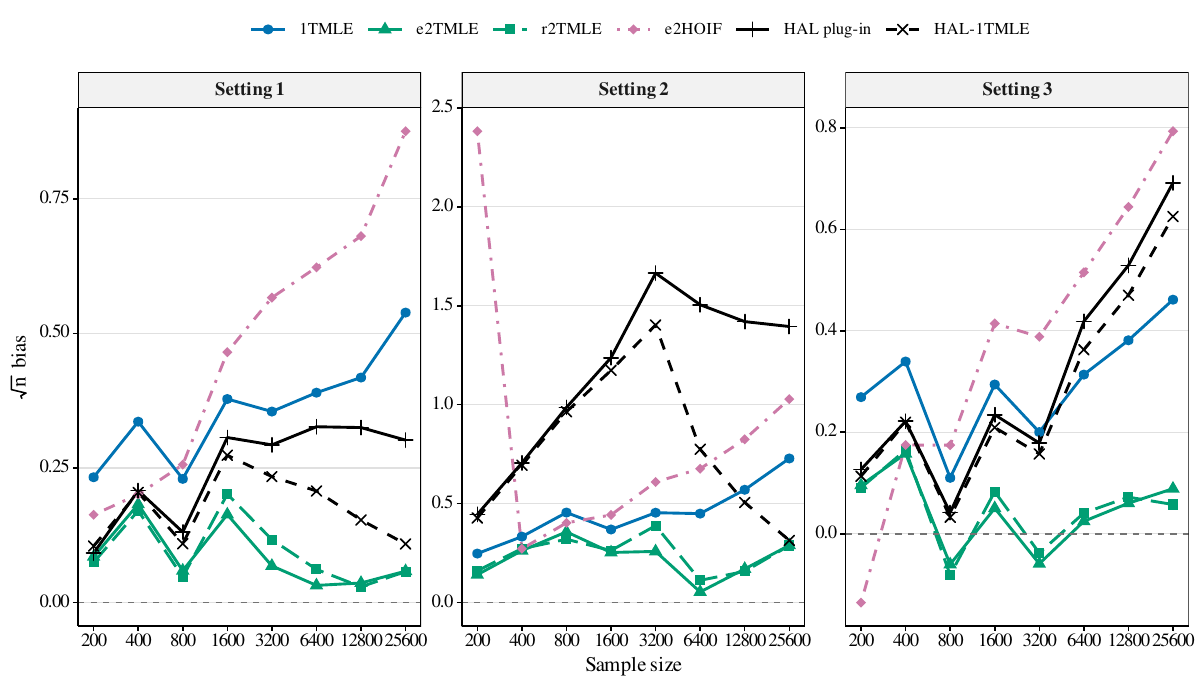}
	\caption{
		Root-$n$ scaled bias for BV0 settings. Settings 1-2: bounded sectional variation with small and large constant variation; Setting 3: increasing sectional variation $K(n)\asymp n^{1/9}$. Nuisance rates are controlled to $n^{-1/6}$ except for 0HAL-1TMLE which uses standard (0-order) HAL initial estimation. 
	}
	\label{fig:cadlag0-rootn-bias-main}
\end{figure}

\begin{figure}[t]
	\centering
	\includegraphics[width=\linewidth]{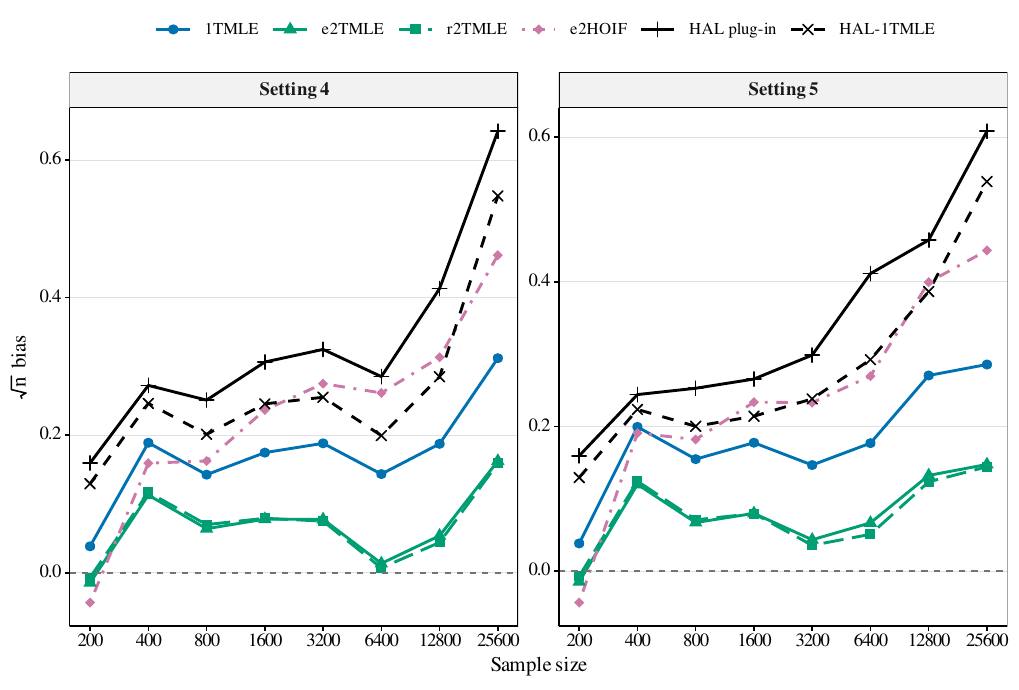}
	\caption{
		Root-$n$ scaled bias for BV1 settings. Settings 4-5: increasing first-order sectional variation with derivative variation $K(n)\asymp n^{3/20}$ and $K(n)\asymp n^{7/30}$, respectively. Nuisance rates are controlled to $n^{-1/6}$ except for 1HAL-1TMLE which uses first-order spline HAL as initial estimation.
	}
	\label{fig:cadlag1-rootn-bias-main}
\end{figure}

Across all settings (Figure \ref{fig:cadlag0-rootn-bias-main} and \ref{fig:cadlag1-rootn-bias-main}), the HAL-based second-order TMLE estimators achieve substantial and stable bias reduction relative to the corresponding first-order TMLE with the same initial nuisance estimators. Both e2TMLE and u2TMLE exhibit comparable root-n-scaled bias performance, without or with undersmoothing, despite relying on nuisance convergence rates as slow as $n^{-1/6}$. These results provide empirical evidence for the theoretical higher-order residual performance in finite samples. 

In Setting 3 with increasing sectional variation norm at rate $K(n) \asymp n^{1/9}$, 1TMLE using either HAL or slowly converging $n^{-1/6}$-rate nuisance estimators shows persistent root-n-scaled bias, whereas 2TMLE remains relatively stable as $n$ increases (Figure \ref{fig:cadlag0-rootn-bias-main}). This performance is consistent with the asymptotic comparison in Table \ref{tab:setting123} and Section \ref{sec:rates}. Similarly, in Settings 4-5 with increasing oscillation through increasing sectional variation norm of first-order derivatives, 2TMLE maintains stable finite-sample bias control and outperforms the state-of-the-art first-order 1HAL-1TMLE competitors (Figure \ref{fig:cadlag1-rootn-bias-main}).

In Setting 1-2 with bounded sectional variation, 0HAL-1TMLE achieves faster theoretical rate of order $n^{-2/3}$; however, 2TMLE exhibits superior finite-sample performance up to $n = 25,600$, particularly in Setting 2 with larger sectional variation constants (Figure \ref{fig:cadlag0-rootn-bias-main}). This suggests that higher-order residual control may dominate the finite-sample performance over a substantial range of practically relevant sample sizes. Thus, even when asymptotic rate improvement is not expected, finite-sample gains may still occur in practice through flexible nuisance estimators and stable third-order residual control.

Although undersmoothing theoretically reduces the regularization bias term $E_n$, the empirical 2TMLE already achieves sufficient control in Settings 1-5, both asymptotically and in finite samples. Therefore, the practical gain from undersmoothing is relatively limited in these scenarios. This demonstrates that the empirical 2TMLE without further undersmoothing may already achieve sufficient regularization bias control in a broad range of nonparametric settings. 

The empirical performance of e2HOIF is substantially less stable. Despite the asymptotic analysis, the finite-sample implementation exhibits dedicate bias-variance trade-off, even when Gram matrix inversion remains numerically stable. The performance may be additionally affected by the discontinuity points that are particularly challenging for global Fourier representations. The theoretical rates such as $k \asymp n^{5/6}/(\log n)^2$ are difficult to implement in practice at moderate sample sizes. Practical tuning must simultaneously balance approximation error, prediction performance, and numerical stability of Gram matrix inversion. These preliminary findings suggest that e2HOIF may require substantially more careful basis selection and optimal-rate-aware tuning than the corresponding HAL-based second-order TMLE estimators. 

\section{Discussion}

In this manuscript, we investigate higher-order estimation constructed based on formal second-order canonical gradients or HAL-based fluctuation parameters. These estimators achieve higher-order residual performance provided proper approximation error control, under related but different nonparametric settings. In addition, we note that all these estimators can be considered estimators of approximate parameters, defined by basis projection, kernel smoothing, or HAL-based fluctuation. 

Under realistic nonparametric settings motivated by complexity and oscillation control through sectional variation, we provide numerical results that empirical second-order estimation can provide meaningful gains not only asymptotically but also in practically relevant finite-sample scenarios. When nuisance estimators converge slowly, the HAL-based second-order TMLE exhibits stable finite-sample improvement over first-order estimators, even under increasing variation and increasing oscillation that deteriorate typical HAL rates. This provides empirical evidence of reliable higher-order residual control and relatively robust performance under different tuning choices. In practice, second-order TMLE may be particularly attractive when flexible nuisance estimators converge slowly, or when large or increasing complexity constants leave persistent finite-sample bias by first-order estimation. 

Our analysis highlights a key conceptual distinction between different higher-order constructions. One challenge arises from lack of second-order pathwise differentiability, which itself requires approximation precision of order $o_P(n^{-1})$ along regular parametric paths. Approximately achieving such precision in large nonparametric models may still be difficult; this partially explains the challenges of HOIF-based estimators in finite samples, despite potentially greater asymptotic gains under corresponding regularity conditions. A different challenge arises from lack of first-order pathwise differentiability of an auxiliary fluctuation parameter, where only $o_P(n^{-1/2})$ precision is required.
Although both constructions lead to additional approximation errors, the resulting practical finite-sample performance may differ substantially. In particular, HAL-based second-order construction addressing the latter difficulty may remain considerably more stable in finite samples while achieving weaker asymptotic guarantees under certain settings. The results also emphasize that asymptotic optimality alone may provide insufficient guidance in realistic finite-sample scenarios for higher-order estimators. 

An important aspect underlying this stability is the score-equation-solving property of HAL-based estimation. HAL solves increasingly rich collections of empirical score equations as the number of active basis grows. This property explains the regularization term control already achieved in empirical 2TMLE, as well as the potential rate improvement through undersmoothing. 
In the meantime, explicit second-order targeting separates approximation from nuisance estimates, avoiding potentially unnecessary variance inflation in aggressive undersmoothing. 
Compared with plug-in estimation solely based on HAL nuisance estimators, the numerical results highlight that explicitly separating nuisance estimation from higher-order construction can provide practical finite-sample advantages, even when asymptotic improvements are not expected, whereas asymptotic rate improvement appears possible with increasing oscillation such as Setting 5. 

The plug-in property of TMLE may be particularly valuable in higher-order estimation. While plug-in estimation is considered a desirable but usually secondary property in first-order semiparametric estimation, it becomes substantially more important when more aggressive debiasing introduces dedicate bias-variance trade-off. By respecting global parameter constraints throughout the first- and second-order targeted updates, the substitution structure of TMLE may partially explain the observed relatively robust finite-sample performance. 

Several limitations and open challenges remain. This work does not systematically investigate positivity challenges that may further affect stability of second-order estimation through increasing instability in clever covariates and canonical gradients. This is particularly important because higher-order covariates can involve more inverse weighting than first-order influence curves, and finite-sample stability is necessary for visible asymptotic gains. Understanding behavior and possible remedies under practical positivity violations warrants an important direction of future work. 

Dimensionality presents practical or theoretical challenges for HAL-based or HOIF-based constructions. Randomized and data-adaptive basis generation may provide favorable computational advantage while largely preserving approximation properties. How different basis growth strategies interact with residual and approximation error controls remains unexplored with the present work. Computational scalability for higher-order estimators remains an important open challenge under high-dimension input with deep interactions. 

The results suggest that classical \holder conditions and Jackson-type minimax rates may not fully utilize realistic complexity assumptions. They may also be affected by discontinuities or other local irregularities, where some basis systems are more sensitive than others. These observations suggest that basis selection remains a practical challenge despite identical minimax rates under ideal conditions, in addition to other critical aspects of bias-variance tuning. One possible direction is combining Lepski's method with cross-validation based tuning and basis selection, with respect to a sequence of fluctuation parameters with decreasing oracle bias. Such procedures would still require simultaneous optimization of approximation precision and Gram matrix stability, which remains  challenging. 

Lastly, the basis representation underlying both empirical second-order HOIF and HAL-based second-order TMLE suggests higher-than-second-order potential that is not fully investigated in this manuscript. Basis representations of fluctuation parameters may substantially simplify the derivation of subsequent canonical gradients, enabling the possibility of partially automated higher-order estimation for maximal asymptotic and finite-sample gains. Such computational approaches may achieve significant practical impact beyond analytically tractable settings.

\clearpage
\appendix

\section{Data Generating Process Details}

We generate i.i.d. copies of $O = (X, A, Y)$ with covariate $X \in [0, 1]^4$, binary treatment $A$, and binary outcome $Y$. 

The covariates $X$ are generated from a rough marginal density using D12/db6 Daubechies wavelet basis with \holder coefficient $\beta_f = 0.1$, with the same Baker-type dependence structure, following \cite{liu2026semiparametricefficientempiricalhigher}. This is for better comparison across methods and to stress test empirical estimator performance. Specifically, 
\begin{align*}
	f_X(t)\propto 1+\exp\set*{
		\frac{1}{2}\sum_{j\in\mathcal J}\sum_\ell
		2^{-j(\beta_f+1/2)}\omega_{j\ell}(t)
		},
	\qquad 0\leq t\leq 1.	
\end{align*}
Here, $\omega_{j\ell}$ denotes the D12/db6 Daubechies wavelet at resolution $j$ and location $\ell$. In the implementation we set $\mathcal J={0,3}$ and the density is normalized numerically.  \texttt{waveslim} R package \citep{whitcher2020package} is used for wavelet evaluations. For Baker-type dependence construction, a pair of independent covariate vectors are generated, and randomly selected coordinate-wise maxima or minima are used as the observation. 

True outcome regression $b_0(x) = E[Y | A = 1, X = x]$ and propensity score $g_0(x) = P(A = 1 | X = x)$ are generated with logits $\eta_b, \eta_g$. We scale probability values to $[0.25, 0.75]$ to ensure positivity. 
\begin{align*}
	b_0(x)
	&=0.25+0.50\cdot\textrm{expit}\{\eta_b(x)\},\\
	g_0(x)
	&=0.25+0.50\cdot\textrm{expit}\{\eta_g(x)\}.
\end{align*}

\subsection{Settings 1-3}

In Setting 1-3, we set 
\begin{align*}
	\eta_b(x)&=\sum_{j=1}^4
	\left\{
		c_b(x_j;a,\kappa)+0.1 \zeta_{bj}h_{0.25}(x_j)
	\right\}, \\
	\eta_g(x)
	&=\sum_{j=1}^4
	\left\{
		c_g(x_j;a,\kappa)+0.1 \zeta_{gj}h_{0.25}(x_j)
	\right\}.
\end{align*}
The mixture component (with wavelets and \holder coefficient 0.25, weighted by 0.1) use the design of Appendix E in \cite{liu2026semiparametricefficientempiricalhigher} with coefficients
\begin{align*}
	\zeta_b&=(-0.2819, 0.4876, -0.1515, -0.1190),\\
	\qquad
	\zeta_g&=(0.09789, 0.08800, -0.4823, 0.4588).
\end{align*}
In addition, $c_b, c_g$ are \cadlag functions piecewise smooth on five sections.  We set
\begin{align*}
	c_b(t;a,\kappa)&=\alpha_r+a \cdot s_r(t;\kappa), \\
	c_g(t;a,\kappa)&=\delta_r+a \cdot u_r(t;\kappa), 	
\end{align*}   
for $t\in I_r, I_r = [0.2(r-1), 0.2r) , r=1,\dots,5$. 
For outcome regression, 
\begin{align*}
\renewcommand{\arraystretch}{1.35}
\begin{array}{c@{\qquad}c@{\qquad}l}
	\hline
	r & \alpha_r & s_r(t;\kappa) \\
	\hline
	1 & -0.05 &
	0.50\sin\{2\kappa\pi(t+1.0)+0.2\}
	+0.20\cos\{3\kappa\pi(t+1.0)-0.1\} \\
	2 & 0.30 &
	0.55\sin\{3\kappa\pi(t+0.6)+0.3\}
	-0.22\cos\{2\kappa\pi(t+0.6)+0.4\} \\
	3 & 0.55 &
	0.60\sin\{2\kappa\pi t+0.15\}
	-0.28\cos\{4\kappa\pi t-0.2\} \\
	4 & 0.25 &
	0.50\sin\{3\kappa\pi(t-0.2)+0.5\}
	+0.18\cos\{2\kappa\pi(t-0.2)+0.3\} \\
	5 & -0.10 &
	0.40\sin\{2\kappa\pi(t-0.6)+0.4\}
	-0.15\cos\{3\kappa\pi(t-0.6)-0.2\} \\
	\hline
\end{array}
\end{align*}
Similarly, for propensity scores, 
\begin{align*}
	\renewcommand{\arraystretch}{1.35}
	\begin{array}{c@{\qquad}c@{\qquad}l}
		\hline
		r & \delta_r & u_r(t;\kappa) \\
		\hline
		1 & -0.90 &
		0.35\sin\{2\kappa\pi(t+1.0)\}
		-0.20\cos\{3\kappa\pi(t+1.0)\} \\
		2 & -0.55 &
		0.45\sin\{3\kappa\pi(t+0.6)\}
		+0.18\cos\{2\kappa\pi(t+0.6)\} \\
		3 & -0.25 &
		0.55\sin\{2\kappa\pi t\}
		-0.22\cos\{4\kappa\pi t\} \\
		4 & -0.45 &
		0.40\sin\{3\kappa\pi(t-0.2)\}
		-0.20\cos\{2\kappa\pi(t-0.2)\} \\
		5 & -0.75 &
		0.30\sin\{2\kappa\pi(t-0.6)\}
		+0.15\cos\{3\kappa\pi(t-0.6)\} \\
		\hline
	\end{array}
\end{align*}
The amplitude parameter $a$ and frequency parameter $\kappa$ are set as
\begin{align*}
	(a,\kappa)=
	\begin{cases}
		(1,3), & \text{Setting 1},\\
		(3,10), & \text{Setting 2},\\
		(1,10(\frac{n}{200})^{2/9}), & \text{Setting 3}.
	\end{cases}
\end{align*}

\subsection{Settings 4-5}

Settings 4-5 use the integrated version of the sectional functions above to control the variation of the first-order derivatives. The derivative sectional variation norms increase at rate $n^{3/20}$ or $n^{7/30}$.  Specifically, 
\begin{align*}
	C_b(t;a,\kappa)&=\int_0^t c_b(u;a,\kappa),du,\\
	\eta_b(x)&=\eta_g(x)=
	\sum_{j=1}^4 C_b(x_j;a, \kappa)
	+ I(x_1>0.5)
	+\sin(\pi x_2).
\end{align*}
The amplitude parameter $a$ and frequency parameter $\kappa$ are set as
\begin{align*}
	(a,\kappa)=
	\begin{cases}
		\left(
		100\left(\frac{n}{200}\right)^{3/40},
		5\left(\frac{n}{200}\right)^{3/40}
		\right),
		& \text{Setting 4},\\[1ex]
		\left(
		100\left(\frac{n}{200}\right)^{7/60},
		5\left(\frac{n}{200}\right)^{7/60}
		\right),
		& \text{Setting 5}.
	\end{cases}
\end{align*}

%\subsection{Controlled Nuisance Errors}
%
%Except for HAL-TMLE competitors, we use the following initial estimator with controlled errors of order $n^{-1/6}$. Estimated probabilities are within $[0.05, 0.95]$. 
%\begin{align*}
%	b &= b_0(x)+n^{-1/6}e_b(x) \\
%	g &= g_0(x)+n^{-1/6}e_g(x). 
%\end{align*}
%The errors are 
%\begin{align*}
%	e_b(x)&=0.05+\{(x_1-0.6)+(x_2-0.6)\}, \\
%	e_g(x)&=0.05+\frac{1}{4}\{(x_3-0.4)+(x_4-0.4)\}. 
%\end{align*}

\clearpage
\bibliography{references.bib}

\end{document}